\documentstyle[12pt]{article}
\begin{document} 

{\bf Preprint SB/F/96-236}
\vskip .5cm
\hrule 
\vskip .5cm
\vskip 1.5cm

\centerline{{\Large BF Topological Theories and 
Infinitely Reducible Systems}}

\vskip.5cm 
\centerline{{\large M. I. Caicedo and A. Restuccia}}
\vskip .5cm
\centerline{\it Universidad Sim\'{o}n Bol\'{\i}var, Departamento de F\'{\i}sica} 
\vskip .5cm
\centerline{\it Apartado postal 89000, Caracas 1080-A, Venezuela.}
\centerline{\it e-mail: mcaicedo@usb.ve,arestu@usb.ve}

\vskip 1cm
{\bf Abstract}
\begin{quotation}
\noindent{\small We present a rigurous disscusion for abelian $BF$ theories in 
which the base manifold of the $U(1)$ bundle is homeomorphic to a Hilbert space.
The theory has an infinte number of stages of reducibility. We specify conditions
on the base manifold under which the covarinat quantization of the system can be
performed unambiguously. Applications of the formulation to the superparticle and 
the supertstring are also discussed.}
\end{quotation}

\vskip 1.5cm  
In this paper we discuss the formulation of a $U(1)$ $BF$ topological  
theory {\cite{re:Blau} formulated over  
an infinite dimensional base manifold (${\cal{M}}_{\infty}$) which is  
locally homeomorphic to a Hilbert space ($\cal{H}$).  
 
There are two main motivations for approaching this problem. The first  
of them is to study such theories as natural generalizations 
of $BF$ theories over finite dimensional manifolds {\cite{re:Blau}} in order 
to investigate their unknown structure of topological observables.  
 
The second motivation for this work is directly related to 
the fact  that $BF$  
actions are infinitely reducible systems, this feature is shared by both the 
10-dimensional
superparticle and the  Type $II$ Green-Schwartz's Superstring. Indeed 
the authors of reference {\cite{re:Steph}} have proposed a 
new  approach for the quantization of  
the Superstring, in this formulation second 
class constraints are absent at the expense of introducing  
an infinite number of first class constraints and stages of 
reducibility. The covariant quantization of infinitely reducible systems  
needs the introduction of an infinite tower of ghosts for ghosts, a 
consistent treatment of this problem is still lacking and therefore one 
finds an obstruction to the covariant quantization of the superstring along 
these lines. Recently another approach to the superstring problem has 
appeared {\cite{re:Bandos}} which makes heavy use of the twistor approach 
but we shall not elaborate on such lines.
 
The main goal of this work is to show that under certain assumptions it is 
possible to give a rigurous treatment of the quantization of infintely
reducible $BF$ systems.   

We consider a $U(1)$ principal bundle over a manifold 
${\cal{M}}_{\infty}(\cong\Re\times\Sigma)$
, which is locally homeomorphic to a Hilbert space $\cal{H}$ 
and let $<\dots{}>$ stand for integration over 
${\cal{M}}_{\infty}$ with an appropiate measure. 
 
Let ${\bf A}^{(\infty)}$ be the $U(1)$ 
connection one form over ${\cal{M}}_{\infty}$ and 
${\bf F}^{(\infty)}={\bf dA}^{(\infty)}$ its 
associated curvature -we well use boldface for all the geometric objects 
in ${\cal{M}}_{\infty}$ and plain text for the geometric objects 
in $\Sigma$-. The $BF$ action is then given by:
 
\begin{equation}
S^{(\infty)}=<{\bf B}\wedge{\bf F}^{(\infty)}>
\label{eq:sgeom} 
\end{equation}
 
\noindent{}where ${\bf B}$ belongs to the dual 
space of two forms. 
In terms of local indices the action can be written as

\begin{equation} 
S^{(\infty)}=<B^{\mu\nu}F_{\mu\nu}^{(\infty)}>
\label{eq:sindex} 
\end{equation}

\noindent{}$B^{\mu\nu}$ being antisymmetric. The convergence on 
the summation is guaranteed from the 
convergence of the expansions of ${\bf B}$ and ${\bf F}^{(\infty)}$ in 
their respective infinite basis. 
 
The canonical analysis of constrained systems and the Dirac
approach to the reduction to the physical submanifold can
be done along the same lines followed in field theory.
Consequently, the action may be rewritten in terms of the
canonical fields:

\begin{equation}
(A_i^{(\infty)},\pi^{i}=2B^{0i})
\label{eq:canfields}
\end{equation}

\noindent{}as a singular action:

\begin{equation} 
S^{(\infty)}=<\pi^i\dot{A}_i^{(\infty)}+A_0^{(\infty)}
\partial_i\pi^i+B^{ij}F_{ij}^{(\infty)}>
\label{eq:scan} 
\end{equation}

\noindent{}where clearly $A_0^{(\infty)}$ and $B^{ij}$ are Lagrange 
multipliers associated to a couple 
of first class constraints whose Poisson bracket algebra is abelian.

As happens in the finite dimensional case, the Gauss constraint, namely: 
$\partial_i\pi^i=0$ is irreducible 
and generates the standard gauge 
transformations on the principal bundle. The remaining 
constraint $F_{ij}^{(\infty)}=0$ brings in the rich topological structure of $BF$ theories
and can be cast in a geometrical way by virtue
of the structure of the basespace ${\cal{M}}_{\infty}$. Indeed, this constraint
is just the statement of flatness of the connection 1-form over $\Sigma$ 
i.e. the vanishing 
of the curvature two form ($F^{(\infty)}=0$). This constraint is reducible
since the Bianchi identity is just:

\begin{equation}
d_2F^{(\infty)}=0
\label{eq:Bianchi}
\end{equation}

\noindent{}where, $d_2$ is the exterior derivative operator acting on $2$ 
forms over $\Sigma$. Obviously, $d_2$ is a reducible operator
 since $d_3d_2=0$, moreover the identity:

\begin{equation}
d_{k+1}d_{k}\equiv{}0
\label{eq:d2}
\end{equation}

\noindent{}shows that the system has an infinite set of 
reducible operators given by the chain: 

\begin{equation} 
{d_1}\rightarrow{d_2}\rightarrow\dots{}\rightarrow{d_N}\rightarrow\dots 
\end{equation}
 
In the finite dimensional case ($dim(\Re\times\Sigma)=1+(D-1))$ 
the $d_{D-1}$ exterior derivative operator acting on $D-1$ forms over $\Sigma$
is trivial, the above chain ends up 
after $D-3$ steps and consequently we are in presence of a system 
with $D-3$ stages of reducibility {\cite{re:Mar-Alv95}}. 
 
We can now formally construct the BRST charge ($\Omega$). 
Following the notation of {\cite{re:Mar-Alv95} it is given by: 
 
\begin{equation} 
\Omega=<C\partial_i\pi^i+
C_1^{ij}F_{ij}+C_{11}^{ijk}\partial_k\mu^1_{ij}+\dots>
\label{eq:Charge1}
\end{equation}
 
\noindent{}an expression which is clearly geometrical: 
 
\begin{equation}
\Omega=<C\partial_i\pi^i+C_1F+C_{11}d\mu^1+C_{111}d\mu^{11}+\dots>
\label{eq:Charge2} 
\end{equation}
 
>From the geometrical point of view $\Omega$ can be thought 
of as the BRST charge corresponding to a 
gauge theory over the minimal sector of the extended (super)phase 
space subjected to the following extra 
set of first class constraints:
 
\begin{eqnarray}
&d\mu^1=0&\nonumber\\
&d\mu^{11}=0&\label{eq:extracons}\\
&d\mu^{[p]}=0,\qquad{}p=3,\dots&
\end{eqnarray}
 
\noindent{}where the bracketed superscript reflects the stage of 
reducibilty to which a particular object belongs. These "extra" 
constraints generate gauge transformations on 
the $C$ fields through the appropiate Poisson brackets. For example: 
 
\begin{equation} 
\delta_{Gauge}C_1(x)=\{C_1(x),<d\mu^1.\varepsilon_1>\}=<d\delta_x\varepsilon_1>=
<\delta_x\delta\varepsilon_1>=\delta\varepsilon_1(x)
\label{eq:Gauge} 
\end{equation}

\noindent{}where: $\delta_x$ is Dirac's delta,
$\delta$ is the co-derivative and $\varepsilon_1$ is a 0-form -the
parameter of the gauge transformation-. According to these definitions
the expression ({\ref{eq:Gauge}}) in components reads as follows: 
 
\begin{equation}
\delta_{Gauge}C_1^{ij}=-\partial_k\varepsilon_1^{kij} 
\end{equation} 
 
The higher level fields have gauge transformations whose parameters
are higher order forms. We find 
 
\begin{eqnarray} 
&\delta_{Gauge}C_{11}(x)=\{C_{11},<d\mu^{11}\varepsilon_{11}>\}=
\delta\theta_{11}(x)\mbox{ or}&\nonumber\\
&\delta_{Gauge}C_{11}^{ijk}(x)=\partial_l\varepsilon_{11}^{lijk}(x)&\nonumber\\
&\dots&\\
&\delta_{Gauge}C_{[p]}(x)=\delta\varepsilon_{[p]}(x)&\nonumber\\
&\dots&\nonumber
\end{eqnarray}
 
At this point it is necessary to go a step backwards in order to 
briefly discuss some issues related with the quantization of
finite dimensional $BF$ actions. As explained before,
when the base manifold of the bundle $({\cal{M}}_D)$ is finite dimensional
the reducibility is up to $D-3$ stages. In reference {\cite{re:Mar-Alv96}} we have shown the gauge 
fixing procedure that explicitly reduces the path integral of any (finitely) 
reducible theory to the physical modes. The process is as follows: one begins by
defining a 
transverse-longitudinal ($T+L$) decomposition with respect to
the highest stage reducibility operator. This decomposition is used in order to 
fix the the longitudinal part of the highest stage auxiliary
fields. After this step one can recursively work backwards stage by stage and end 
up with an effective action and path integral
formulated in terms of the transverse (physical) degrees of freedom only.
The $T+L$ decomposition just described
defines the set of admissible gauge fixing conditions for the auxiliary fields 
associated to reducible constraints
through the fixing of the longitudinal part of such fields.

 In the case under discussion ($BF$ theories) the $T+L$ decomposition does not
break the $SO(D-1)$ global invariance and at the end one may 
recover a covariant effective action. For the sake of completeness we will now
present the procedure for a 4-dimensional $BF$ theory. In four dimensions the highest
reducibility operator is given by the exterior derivative acting on three forms (or their duals the zero forms) this suggests the following $T+L$ decomposition for one forms:

\begin{eqnarray}
&v_l=\partial{}v+v_l^{T}&\mbox{  where}\\
&\partial_lv_l^{T}=0&
\label{eq:TL}
\end{eqnarray}

\noindent{}according to this formula, the gauge transformations for the 
ghost fields can be written as

\begin{eqnarray}
&\delta_{Gauge}C_{1}^{j}(x)=-\epsilon^{jk}\partial\varepsilon_{1k}^{T}(x)&\\
&\delta_{Gauge}C_{11}^{jk}(x)=-\epsilon^{ijk}\partial_i\varepsilon_{11}(x)&
\end{eqnarray}

\noindent{}where $\varepsilon_1$, $\varepsilon_{11}$... are appropiate forms.

It is convenient to introduce dual objects to the ghosts 
($C_{11}^{ijk}\equiv\epsilon^{ijkl}K_l$
and $C_1^{ij}\equiv\epsilon^{ijkl}K_{kl}$) since in such terms the gauge 
transformations can be cast as follows

\begin{eqnarray}
&\delta_{Gauge}K_l(x)=\partial_l\varepsilon&\\
&\delta_{Gauge}K_{kl}(x)=\partial_{[k}\varepsilon^{T}_{l]}&
\end{eqnarray}

\noindent{}showing that it is possible to completely fix the gauge for the 
longitudinal components of $K_l$ and $K_{kl}$ without breaking the 
explicit $SO(D-1)$ global invariance. With little extra effort it
 is possible to rearrange the full set of fields (including
the tower of ghosts for ghosts) in order to end up with 
a manifestly covariant $SO(D)$ effective action ($-S_{eff}^{(D)}$) whose 
exponential we functionally integrate in all the fields 
-with unit Liouville measure ($D\mu$)- in order to 
find a covariant path integral:

\begin{equation}
I^{(D)}_{Cov}=\int{}D\mu{}exp^{-S_{eff}^{(D)}}
\end{equation}

The covariant gauge fixing is admissible and as a consequence of this we 
may recall the $BFV$ theorem to show that the result of the last path 
integral
 coincides with another gauge fixed path integral within the admissible set. 

After the above discussion the problem when treating the infinite dimensional 
case becomes obvious, there is not such a $T+L$ decomposition so it seems that 
the programme just outlined will not be applicable, nevertheless 
we shall see that by adequately complementing the conditions given for 
the base manifold one can rigurously treat the quantization of the infinite 
dimensional $BF$ theories.

Before embarking on such entrerpise we will see how the light cone
gauge technology can be used in order to build an effective action an path 
integral over the physical modes only. The main idea here consists in 
remembering that in order to build a correct quantum theory it is enough to
 find the right set of physical modes even at the expense of loosing
 the manifiest covariance. In the light cone gauge we can truncate the infinite
 tower of ghosts and their descendants at any stage of reducibility  by 
completely 
breaking the gauge invariance. As an explicit example let us consider such
 breaking at the very lowest (zero) stage of reducibility. We begin by 
chosing appropiate coordinates i.e. $\{0,i\}\rightarrow\{-,+,I\}$ and 
use $x^-$ as "time", the gauge transformations for the first stage 
ghosts are then given by

\begin{eqnarray}
&\delta_{Gauge}C_1^{+I}=-\partial_K\varepsilon_1^{K+J}&\\
&\delta_{Gauge}C_1^{IJ}=
-\partial_+\varepsilon_1^{+IJ}-\partial_K\varepsilon_1^{KIJ}&
\end{eqnarray}

The reducibility of the system manifests itself in the existence of the 
following residual gauge transformations

\begin{equation}
\varepsilon_1^{KIJ}\rightarrow\varepsilon_1^{KIJ}+\partial_P\varepsilon_{11}^{PKIJ}
\end{equation}

\noindent{}which is nothing but the coordinate expression of 
$dd\varepsilon_{11}=0$.

If we assume the invertibility of $\partial_+$ -as is usually done when 
using the LCG- we can use the gauge fixing condition

\begin{equation}
C_1^{IJ}=0
\end{equation}

\noindent{}leaving $C_1^{+J}$ as independent fields, these last fields are 
invariant
 under the residual gauge. Moreover, the constraints that generate the 
residual gauge, namely:

\begin{equation}
\partial_{[k}\mu_{ij]}^1=0
\end{equation}

\noindent{}can also be explicitly solved for $\mu^1_{IJ}$. Indeed, using 
the LCG we get

\begin{equation}
\partial_+\mu_{IJ}^1+\partial_J\mu_{+I}^1+\partial_I\mu_{J+}^1=0
\end{equation}

from which one obtains:

\begin{equation}
\mu^1_{IJ}=-\frac{1}{\partial_+}[\partial_J\mu^1_{+I}+\partial^1_{J+}]
\end{equation}

\noindent{}notice that this solution implies the identity:

\begin{equation}
\partial_{[I}\mu^1_{JK]}=0
\end{equation}

Since we have eliminated conjugate pairs we have shown an explicit
canonical reduction to the finite degrees of freedom of 
the theory. We may now 
approach the construction of the light con gauge path integral over
 the physical modes by functional integrating the exponential of the
 light cone action over the the independ
ent variables with unit measure ($D\mu_{Phys}$) as usual:

\begin{equation}
I^{(\infty)}_{LCG}=\int{}D\mu_{Phys}{}exp^{-S_{LCG}^{(\infty)}}
\end{equation}

We will now turn our attention back to the original problem: 
formulating a covariant effective action and path integral for 
the infinite dimensional $BF$ theory whose classical action is 
$S^{(\infty)}$. We will begin our detailed analysis by 
assuming the existtence of a sucesion of
finite ($D$-dimensional) manifolds ${\cal{M}}_D$ such that 
$M_{\infty}$ is contractible to all ${\cal{M}}_D$ 
for big enough $D$. Let us now consider the sucesion of $U(1)$ $BF$ 
theories over ${\cal{M}}_D$ and their associated reducible constraints

\begin{equation}
F^{(D)}=0
\label{eq:FD}
\end{equation}

The contractibility of ${\cal{M}}_{\infty}$ to ${\cal{M}}_D$ 
implies that the $U(1)$ principal bundles over such manifolds 
are equivalent. Since flat connections over one principal bundle have 
associated flat connections over the equivalent
bundle there is a one to one correspondence between the space of solutions
of ({\ref{eq:FD}}) and  those of 

\begin{equation}
F^{(\infty)}=0
\label{eq:Finf}
\end{equation}

We have then shown that under our assumptions both theories (finite 
dimensional and infinite dimensional) have the same number of degrees 
of freedom. We will take advantage of this fact in order to build the 
quantum theory for the infinite dimensional $BF$ theory.

To begin with we  first notice that under the assumptions, the finite dimensional 
LCG effective action has a well defined limit given by: 

\begin{equation}
\lim_{D\rightarrow\infty}S_{eff(LCG)}^{(D)}=S_{eff(LCG)}^{(\infty)}
\end{equation}

on the other hand, and by virtue of the BFV theorem the finite dimensional 
LCG and covariant path integrals are equivalent i.e.

\begin{equation}
I^{(D)}_{LCG}\cong{}I^{(D)}_{Cov}
\end{equation}

This allows the following definition for the infinite dimensional 
covariant $BF$ path integral:

\begin{equation}
\lim_{D\rightarrow\infty}I^{(D)}_{Cov}\equiv{}I^{(\infty)}_{Cov}
\end{equation}

The argument we have elaborated shows a way to solve the problem of infinetely
 reducible systems. The formulation of the problem needs additional
 structure contained in our assumption of contractibility of
 ${\cal{M}}_{\infty}$ to ${\cal{M}}_D$. {\it If this st
abilizing property can be added to a system with infinite stages 
of reducibility there is an unambiguous solution to the problem.}

If we compare ({\ref{eq:Bianchi}}) and ({\ref{eq:d2}}) with the 
reducibility problem found for the superparticle {\cite{re:Steph}}
 or the superstring  we find a close similarity.
 In fact, for the superparticle we have that for 
the reducible first class constraints

\begin{eqnarray}
&\gamma^\mu{}p_\mu\psi=0&\\
&(\gamma^\mu{}p_\mu)(\gamma^\rho{}p_\rho)=0
\end{eqnarray}

If we are able to introduce -in a consistent way- an odd weight for the 
operator $\gamma^\mu{}p_\mu$ 
acting over odd elements of an infinite dimensional Grassman algebra
${\cal G}_{\infty}$ we may expect the set ${\cal{M}}_D$  
to be related with a sucesion of finite dimensional Grassman algebras
 ${\cal G}_{(D)}$ approaching  ${\cal G}_{\infty}$
 when $D\rightarrow\infty$.

\vskip .5cm
{\bf Akcnowledgments}
\vskip .5cm
This work has been partially supported by 
the "Decanato de Investigaciones de la Universidad 
Sim\'on Bol\'{\i}var" through the research fund number S10-CB-812.

\end{document}